\documentclass[aps,prl,twocolumn,showpacs,floatfix,nobibnotes,superscriptaddress,longbibliography,nofootinbib]{revtex4-1}

\usepackage[utf8]{inputenc}
\usepackage{amssymb}
\usepackage{graphicx}
\usepackage{verbatim}
\usepackage{epsfig}
\usepackage{bm}
\usepackage{color}
\usepackage{xcolor}
\usepackage{float}
\usepackage{dcolumn}
\usepackage{multirow} 
\usepackage{physics}
\usepackage{braket}
\usepackage{ytableau}
\usepackage{footnote}
\usepackage{amsmath}
\usepackage[unicode=true,
  linktocpage,
  linkbordercolor={0.5 0.5 1},
  citebordercolor={0.5 1 0.5},
  colorlinks=true,
  linkcolor=blue,
  citecolor=blue,
  urlcolor=blue]{hyperref}
\usepackage{soul}
\setstcolor{purple}
\soulregister{\cite}{7}

\begin{document}
%----------------------------------------
%----------------TITLE-------------------
%----------------------------------------
\title{Role of spin-isospin symmetries in nuclear \texorpdfstring{$\beta$}{β}-decays}

%----------------------------------------
%---------------AUTHORS------------------
%----------------------------------------
\author{Simone Salvatore Li Muli}
\email{simone.limuli@chalmers.se}
\affiliation{Department of Physics, Chalmers University of Technology, SE-412 96 G\"oteborg, Sweden}
 
\author{Tor R. Djärv}
\email{djarvtr@ornl.gov}
\affiliation{Computing and Computational Sciences Directorate, Oak Ridge National Laboratory, USA}

\author{Christian Forssén}
\email{christian.forssen@chalmers.se}
\affiliation{Department of Physics, Chalmers University of Technology, SE-412 96 G\"oteborg, Sweden}

\author{Daniel R. Phillips}
\email{phillid1@ohio.edu}
\affiliation{Department of Physics and Astronomy and Institute of Nuclear and Particle Physics, Ohio University, Athens, Ohio 45701, USA}

\affiliation{Department of Physics, Chalmers University of Technology, SE-412 96 G\"oteborg, Sweden}

%----------------------------------------
%---------------ABSTRACT-----------------
%----------------------------------------
\begin{abstract}
A century ago, Wigner's SU(4) symmetry was introduced to explain the properties of atomic nuclei. 
Despite recent revived interest, its impact on nuclear structure, transitions, and reactions has not been fully explored.
Here, we show that a variety of high-fidelity nuclear interactions predict nuclear states that have $\geq 90$\% probability of being in a single SU(4) irreducible representation.
Meanwhile, our analysis of axial current operators in chiral effective field theory ($\chi$EFT) reveals that one-body currents at low momentum transfer act only within SU(4) irreducible representations, while two-body currents connect different representations.
These selection rules interfere with the expected convergence pattern of the  $\chi$EFT expansion and explain key phenomenological observations, e.g., the unnaturally large two-body corrections to the axial-current matrix elements in eight-body nuclei.
\end{abstract}

\maketitle
%----------------------------------------
%-------------INTRODUCTION---------------
%----------------------------------------
\textit{Introduction}---\footnote{{\emph{
Notice: This manuscript has been authored, in part, by UT-Battelle, LLC, under contract DE-AC05-00OR22725 with the US Department of Energy (DOE). The US government retains and the publisher, by accepting the article for publication, acknowledges that the US government retains a nonexclusive, paid-up, irrevocable, worldwide license to publish or reproduce the published form of this manuscript, or allow others to do so, for US government purposes. DOE will provide public access to these results of federally sponsored research in accordance with the DOE Public Access Plan (https://www.energy.gov/doe-public-access-plan).}}}Nuclear $\beta$-decay transmutes atoms of one element into atoms of a different element. Over the last decades, first-principles calculations of $\beta$-decay have deepened our understanding of this low-energy manifestation of the Standard Model's weak interactions~\cite{Nav97,Sch02,Mar11,Eks14,Pas17,Gys19,Kin20,Sar21,Gli22,Kin23,Lon24,Gen25}. In Ref.~\cite{Gys19}, \emph{ab initio} many-body calculations %and $\beta$-decay operators 
using a low-energy effective field theory of QCD explained observed $\beta$-decay rates in elements from tritium to tin---without the need for the empirical quenching factors employed in the nuclear shell model. Recent {\it ab initio} calculations like these facilitate the use of precise measurements of $\beta$-decay as an arena for tests of the Standard Model~\cite{Sev06,Gon19,Cir19,Bur22,Bro23,Cir24A,Cir24B}.

In this letter we show that the pattern of $\beta$-decay matrix elements seen in these forefront calculations is driven by a symmetry identified in one of the earliest attempts to understand the systematics of nuclear structure: Wigner's SU(4) spin-isospin symmetry~\cite{Wig37,Wig39}. Interest in SU(4) symmetry (from now on we label Wigner's SU(4) as SU(4) for brevity) in light nuclei was revived in the mid-1990s, due to the observation that a (contracted) SU(4) symmetry of nuclear forces emerges in the large-$N_c$ (large number of colors) limit of QCD~\cite{Kap96,Cor08,Phi13}. More recently, lattice QCD calculations at the SU(3)-flavor symmetric evinced SU(6) symmetry in baryon-baryon scattering amplitudes~\cite{Wag17}.
Lee {\it et al.} showed that chiral EFT ($\chi$EFT) potentials have SU(4) symmetry if evaluated at a particular resolution scale, and that this symmetry is visible in matrix elements of the nucleon-nucleon (NN) interaction in ${}^{30}$P~\cite{Lee21}. Moreover, the LO interactions in an EFT for nuclei in which pions have been integrated out are SU(4) invariant~\cite{Meh99} and Lin {\it et al.} recently demonstrated that this explains why certain amplitudes for the M1 threshold capture of photons in the $A=3$ system are vanishingly small~\cite{Lin22}. Such suppressions are not predicted by the power counting of $\chi$EFT, because SU(4) is an emergent symmetry in that EFT.

The relevance of SU(4) to $\beta$-decay stems from a long-established consequence of the group's structure: the Gamow-Teller operator, being a generator of the SU(4) algebra, cannot induce transitions between different irreducible representations (irreps)~\cite{Wig39,Hec69,Muk74,Vog93}. While this selection rule is well-known, its physical utility hinges on the degree to which nuclei computed from first principles are themselves pure representations of SU(4) symmetric states. To investigate this, we performed \textit{ab initio} No-Core Shell Model (NCSM) calculations for nuclei with $A=3-8$. Following a strategy similar to earlier shell-model work in the sd-shell~\cite{Vog93}, we decomposed our resulting wave functions into SU(4) irreps, using an efficient Lanczos-based algorithm well-suited to large model spaces~\cite{Joh15}. 

Our novel finding is that {\it ab initio} wave functions for light nuclei based on a number of sophisticated $\chi$EFT forces are strongly aligned with a single SU(4) irrep.  
This holds for $\chi$EFT forces constructed at leading order (LO), as well as NLO and N2LO in the $\chi$EFT expansion. It also holds under a variety of fitting strategies, and regardless of whether a three-nucleon (3N) force is included in the calculation or not. 
Moreover, our analysis of the axial-current operator derived from $\chi$EFT reveals a specific pattern: while the dominant one-body part of the current is largely block-diagonal in SU(4) irreps, two-body currents do not share this property. SU(4) symmetry thus imposes approximate selection rules on low-energy electroweak processes.

%-----------------------------------------
%---------------THEORY-------------------
%----------------------------------------
\textit{The SU(4) group and its irreducible representations}---To introduce our original work we first give a summary of relevant results obtained already by many authors, see for instance Refs.~\cite{Wig37,Wig39,Hec69,Rad74,Tal93_book,Nan72,Muk74,Vog93}. 
SU(4) symmetry is invariance under the group of special unitary transformations in spin-isospin space. The fifteen group generators are:
$T_a = \frac{1}{2} \ \sum_k \ \tau^{(k)}_a$, 
$S_b = \frac{1}{2} \ \sum_k \ \sigma^{(k)}_b$, and 
$G_{ab} = \frac{1}{2} \ \sum_k \ \tau^{(k)}_a \ \sigma^{(k)}_b$,
with $a,b=1,2,3$, and $k$ a single-particle index.
The quadratic Casimir, $C_2$, commutes with all these generators and so is invariant under SU(4) transformations:
\begin{eqnarray}
C_2 = T^a \ T_a + S^b \ S_b + G^{ab} \ G_{ab} \label{eq: casimir} \ ,
\end{eqnarray}
where Einstein convention of summing repeated indices is used.
States populating identical irreps of SU(4) have the same expectation value of $C_2$.  
They also belong to the same irreps of the permutation group and so can be classified using Young tableaux. In Table~\ref{tab:1} in the End Matter we list the irreps of SU(4) with the lowest values for the quadratic Casimir operator in systems of $A=2-8$ nucleons. We identify them by their degeneracy $[\mathbf n]$, the value of the quadratic Casimir $\braket{C_2}$, and the Young tableau specifying the permutational symmetry.

\textit{SU(4) structure of nuclear wave functions}---We seek to expand a nuclear state $\ket{\Psi_{JM_T}}$, with good total angular momentum $J$ and isospin projection $M_T$, in terms of irreps of the SU(4) group. Our representation of SU(4) states is based on Ref.~\cite{Hec69}; key results from this reference are summarized in the End Matter for convenience. The basis states are constructed as the tensor product of a many-body spatial state $\ket{[\tilde f];\alpha L M_L}$, and a coupled spin-isospin state $\ket{[f];STM_SM_T}$. Here, $[f]$ is a Young tableaux with dual tableaux $[\tilde f]$, $L$, $S$ and $T$ are the total orbital, spin and isospin quantum numbers with projections $M_L, M_S$ and $M_T$ respectively, while $\alpha$ includes spatial quantum numbers beyond $L$ and $M_L$. The states $\ket{[f]; S T M_S M_T; \alpha L M_L} \equiv \ket{[f]; S T M_S  M_T} \otimes \ket{[\tilde f]; \alpha LM_L}$ form a complete basis; nuclear states are expanded as
\begin{eqnarray}
     \ket{\Psi_{JM_T}}&=& \sum_{[f]} \sum_{\substack{\alpha LM_L \\ S M_S T}} \braket{[f]; S T M_S M_T; \alpha LM_L|\Psi_{JM_T}}  \nonumber \\ && \ket{[f]; S T M_S M_T; \alpha L M_L}  \ . \label{eq: expansion}
\end{eqnarray}
To calculate the probability, $P([f])$, of populating a specific irrep $[f]$ one needs to evaluate several scalar products of $\ket{\Psi_{JM_T}}$ with eigenstates of the Casimir.
Since the Casimir operator is independent of the nucleons' orbital degrees of freedom, diagonalization in terms of nuclear many-body states increases the degeneracy of the operator factorially. Explicit decomposition becomes prohibitively expensive in large model spaces, like those used in many {\it ab initio} techniques. 

The Lanczos algorithm offers an alternative and efficient way to achieve this decomposition \cite{Whi80}.  This method has been applied to investigate nuclear properties, such as the spin-orbit decomposition of nuclear states \cite{Joh15} and spatial symmetries \cite{Gue00,Mcc20,Mcc24}. In this approach, we start the Lanczos iteration to diagonalize the quadratic Casimir using $\ket{\Psi_{JM_T}}$ as the pivot vector $\ket{K_1}$. The Lanczos algorithm then generates a Krylov space $\{ \ket{\Psi_{JM_T}}, \ket{K_j}; j=2, 3, \ldots \}$ and diagonalization of the tridiagonal matrix that is the representation of $C_2$ in this space yields the eigenvalues of $C_2$, as well as its eigenvectors, expressed as linear combinations of the Krylov vectors. This allows us to simply read off the amplitude for $\ket{\Psi_{JM_T}}$ to be in any of the states with eigenvalue $\braket{C_2}$. Notably, this decomposition accumulates the probabilities of populating irreps which share the same value for $\braket{C_2}$. This feature does not create problems in our study, since all the analyzed irreps have distinct values for $\braket{C_2}$, see Table~\ref{tab:1} in the End Matter.

\textit{SU(4) structure of nuclear axial currents}---Following Ref.~\cite{Bar18}, the LO axial-current operator in $\chi$EFT is
\begin{eqnarray}
    j^{\mathrm{LO}}_{5,a,b}(\mathbf q) = - \frac{g_A}{2}\sum_k \ e^{i \mathbf q\cdot \mathbf r^{(k)}} \ \tau^{(k)}_a \ \sigma^{(k)}_b \ , \label{eq: lo_cheft_axial_current}
\end{eqnarray}
where $g_A$ is the nucleon axial coupling constant, and $k$ is a nucleon label.
Up to and including N3LO (order $Q^3$ relative to leading) in the $\chi$EFT expansion, the only one-body correction to Eq.~(\ref{eq: lo_cheft_axial_current}) is a relativistic effect with a pre-factor of $1/(8 M_N^2)$ and a size of $< 1\%$ relative to LO.
The operator in Eq.~(\ref{eq: lo_cheft_axial_current}) transforms as a $[211]$ tensor under SU(4). This structure connects states populating different SU(4) irreps, as is shown explicitly for a selection of transitions in the section \textit{SU(4) tensor products} of the End Matter. However, in the long-wavelength limit the current reduces to the $A$-body Gamow-Teller operator; a special case of a $[211]$ tensor without orbital structure. Being a generator of SU(4), the Gamow-Teller operator is unable to transition states to different irreps.
The approximate SU(4) symmetry of nuclear states therefore leads to approximate selection rules for transitions mediated by the one-body operator in the ${\bf q} \rightarrow 0$ limit.

Two-body currents also appear in the $\chi$EFT expansion of the nuclear axial current, and enter at N2LO/N3LO in the version of $\chi$EFT with/without an explicit $\Delta(1232)$. The complete current up to N3LO can be found in Eqs.~(2.5)--(2.11) of Ref.~\cite{Bar18} and includes both a short-distance piece and a (spatial) tensor structure. 
The spin-isospin structure of the dominant pieces of the two-body current can be reexpressed in terms of two-body SU(4) generators, see Eqs.~(\ref{eq: 2b_axial_current_1},\ref{eq: 2b_axial_current_2}) in the End Matter.
As a result, two-body axial current operators are block diagonal in SU(4) irreps of two-particle states. They are not, however, block diagonal in the $A$-body SU(4) irreps.
Thus, while SU(4) symmetry may still generate certain selection rules for two-body transitions, their impact must be studied on a case-by-case basis in $A$-body systems. 
In the section \textit{SU(4) tensor products} of the End Matter we show a selection of SU(4) transitions induced by two-body currents. In future work we will decompose $\chi$EFT two-body axial currents into SU(4) tensors, thereby yielding 
more specific selection rules.

%-----------------------------------------
%---------------METHODS-------------------
%----------------------------------------
\textit{Methods}---Weak operators and nuclear potentials are calculated using $\chi$EFT. 
In this framework, nuclear operators are expanded in powers of the small parameter $Q \equiv \max({m_\pi,p})/{\Lambda_b}$, where $m_\pi$ is the pion mass, $p$ is the characteristic momentum of nucleons in the nucleus, and $\Lambda_b \approx 500-600$ MeV is the breakdown momentum of the EFT~\cite{Mel17,Wes21}. 
Previously we discussed the axial current up to relative order $Q^3$ in this expansion. 
To construct nuclear Hamiltonians at the same order, N2LO potentials are needed; these include both NN and 3N forces. $\chi$EFT potentials must be regularized and fit to data. Both aspects of this procedure produce variability in predictions---even among potentials that are nominally of the same $\chi$EFT order.   

We solve the many-body Schrödinger equation using the NCSM~\cite{Bar13}. In the NCSM, the Hamiltonian is projected on a harmonic-oscillator basis and becomes a large but sparse matrix. The diagonalization is performed iteratively using the Lanczos algorithm. 
The NCSM includes all nucleons as dynamical degrees of freedom, but the full Hilbert space is truncated on total harmonic-oscillator excitations, 
parameterized by $N_{\mathrm{max}}$.
Here we employ the pAntoine~\cite{Cau99,Nav04,For17} and JupiterNCSM~\cite{Dja22} codes. The latter implementation handles full 3N forces.

%----------------------------------------
%--------------RESULTS-------------------
%----------------------------------------
\begin{figure*}[ht]
    \centering
    \includegraphics[scale=1]{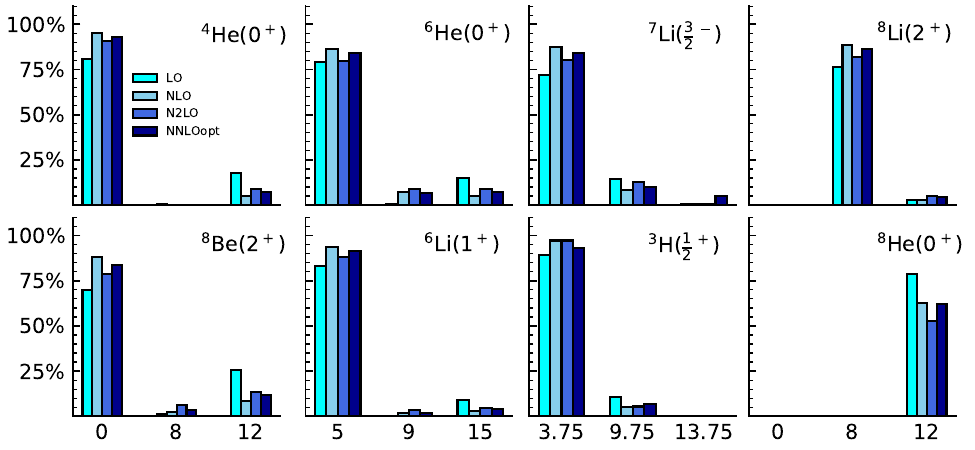}
\caption{Probabilities of selected nuclei to populate SU(4) irreps. The irreps are identified by the corresponding value of the quadratic Casimir, see Table~\ref{tab:1}. Nuclear wave functions are obtained from $\chi$EFT interactions at LO (cyan), NLO (light-blue), N2LO (blue) \cite{Wes21}. Results for the NNLOopt~\cite{Eks13} interaction---which does not include three-body forces---are in dark blue.}
\label{fig:1} 
\end{figure*}

\textit{Results}---We use the Lanczos method to decompose nuclear states for $A=3-8$ in SU(4) irreps. The probabilities, $P([f])$, are shown in Fig.~\ref{fig:1}, where we label SU(4) irreps by the corresponding value of $C_2$. 
Despite the presence of explicit SU(4)-symmetry-breaking in the Coulomb and nuclear forces---and different amounts of breaking in different force implementations---the decompositions of nuclear eigenstates are largely insensitive to the choice of nuclear Hamiltonian and chiral order. We studied the sequence of LO (cyan), NLO (light-blue), N2LO (blue) $\chi$EFT interactions from Ref.~\cite{Wes21} and the NNLOopt interaction without 3N forces (dark-blue) from Ref.~\cite{Eks13}. All produce essentially the same SU(4) decompositions. The main components in the SU(4) decomposition also converge quickly with $N_{\rm max}$.
Nuclear states for $A=3-8$ have more than 90\% of their normalization in a specific SU(4) irrep.
When 3N forces are included, as in the N2LO interaction from Ref.~\cite{Wes21}, the probability of the main component decreases slightly, indicating that 3N operators produce stronger symmetry breaking. 

We use our results to explain the $\beta$-decay patterns noted in \textit{ab initio} calculations, see for instance Ref.~\cite{Kin20}. Gamow-Teller matrix elements can be evaluated analytically in the SU(4) limit \cite{Hec69}, where nuclear states have $L$, $S$, and $T$ as additional good quantum numbers. Following the approach described in the End Matter, we obtain
\begin{eqnarray}
&&\braket{\Psi'_{\text{SU(4)}}|GT_\pm|\Psi_{\text{SU(4)}}}= \nonumber \\ 
&& (-1)^{L+J+S'+1} \ \sqrt{\ 8 \ (2J+1)\ (2S'+1) \ C_2([f])} \label{eq: matrix_element_exact_su4} \\ 
&&
C_{JM,10}^{J'M} C_{TM_T,1\pm1}^{T'M_T'} W_{[f]ST,[211]11}^{[f']S'T'} \begin{Bmatrix}
S & L & J \\
J' & 1 & S'
\end{Bmatrix} \delta_{[f'][f]}\ , \nonumber
\end{eqnarray}
where $W_{[f]ST,[211]11}^{[f']S'T'}$ are reduced Wigner coefficients that generalize Clebsch-Gordan coefficients to the coupling of SU(4) states, $C_2([f])$ is the quadratic Casimir operator of SU(4) evaluated in the $[f]$ irrep, and the last two terms are a Wigner $6$-$j$ coefficient and a Kronecker delta on the SU(4) irreps.

In Table~\ref{tab:2}, we compare matrix elements evaluated using Eq.~(\ref{eq: matrix_element_exact_su4}) to \textit{ab initio} NCSM calculations using LO, NLO and NNLOopt $\chi$EFT interactions and the variational Monte Carlo (VMC) calculations of Ref.~\cite{Kin20}. For SU(4)-allowed transitions two observations are noteworthy:
First, Eq.~(\ref{eq: matrix_element_exact_su4}) gives results remarkably close to the \textit{ab initio} computations---the relative difference for these transitions is less than 10\%, irrespective of the interaction. Second, the Gamow-Teller matrix element is very stable with respect to the order of the $\chi$EFT interaction used. For example, in ${}^6$He the result with a leading-order $\chi$EFT wave function is within 4\% of the results at $O(Q^3)$ (albeit without three-body forces). This happens because the matrix element is symmetry protected.
The emergence of approximate SU(4) symmetry means that all models yield matrix elements for Gamow-Teller operators that are close to the prediction of Eq.~(\ref{eq: matrix_element_exact_su4}).

Furthermore, our findings also explain the suppression of Gamow-Teller matrix elements in $A=8$ nuclei observed in Refs.~\cite{Gys19,Kin20}, because these transitions are SU(4) forbidden. While the wave function of $^8$Be(2$^+$) mostly populates the [000] irrep, see Fig.~\ref{fig:1}, the two isobaric partners $^8$Li(2$^+$) and $^8$B(2$^+$) have zero overlap with this representation. The decays $^8$Li(2$^+$)$\rightarrow ^8$Be(2$^+$) and $^8$B(2$^+$)$\rightarrow ^8$Be(2$^+$) primarily occur within the irrep [211] of SU(4) and are suppressed due to the small weight of this component in $^8$Be(2$^+$). 
Similar arguments apply to $^8$He(0$^+$)$\rightarrow ^8$Li(1$^+$), with the exception that this transition occurs mostly in the irrep [220]. 

This elucidates the reason for the large two-body corrections ($\approx 30\%$) observed in the nuclear matrix elements of $A=8$ decays \cite{Kin20}; SU(4) symmetry suppresses one-body, but not two-body matrix elements. 
In fact a component of the spin-isospin two-body axial current---which is linear in the two-body generator $G^{(2)}_{ab}$, see Eq.~(\ref{eq: 2b_axial_current_1}) in the End Matter---has a tensor structure $[211]$. The dominant component of the wavefunction of $^8$Be(2$^+$) has a tensor structure $[000]$, while the dominant parts of $^8$Li(2$^+$) and $^8$B(2$^+$) wave functions have structure $[211]$. These large components are thus connected by two-body axial currents, resulting in a non-suppressed contribution.

\begingroup
\begin{table*}[htb]
\centering
\begin{tabular}{c c c cccc c}
\toprule
\multicolumn{1}{c}{\multirow{2}{*}{Transition}} & Within & \multirow{2}{*}{Eq.~(\ref{eq: matrix_element_exact_su4})} & \multicolumn{4}{c}{This work (NCSM)} & VMC \\
& irrep & & LO & NLO & NNLOopt & $N_\mathrm{max}$ & \cite{Kin20} \\
\hline
\hspace{0.8em}$^3$H$\bigl ( \frac{1}{2}^+\bigr )$ $\longrightarrow {}^3$He$\bigl ( \frac{1}{2}^+\bigr )$ & \checkmark & 2.449 & 2.267 & 2.332 & 2.313 & 18 & --\\
$^6$He(0$^+$)  $\longrightarrow {}^6$Li(1$^+$) & \checkmark & 2.449 & 2.178 & 2.259 & 2.260 & 14 & 2.200\\
$^7$Be$\bigl ( \frac{3}{2}^-\bigr )$ $\longrightarrow {}^7$Li$\bigl ( \frac{3}{2}^-\bigr )$ & \checkmark & 2.582 & 2.301 & 2.375 & 2.357 & 12 & 2.317\\
$^7$Be$\bigl ( \frac{3}{2}^-\bigr )$ $\longrightarrow {}^7$Li$\bigl ( \frac{1}{2}^-\bigr )$ & \checkmark & 2.309 & 2.086 & 2.178 & 2.175 & 12 & 2.157\\
\hspace{0.6em}$^8$Li(2$^+$) $\longrightarrow {}^8$Be(2$^+$) & \text{\sffamily X} & 0.0 & 0.018 & 0.081 & 0.093 & 10 & 0.147\\
$^8$He(0$^+$) $\longrightarrow {}^8$Li(1$^+$) & \text{\sffamily X} & 0.0 & 0.066 & 0.364 & 0.335 & 10 &{0.386}
\end{tabular}
\caption{Reduced Gamow-Teller matrix elements obtained with Eq.~(\ref{eq: matrix_element_exact_su4}), compared to several \textit{ab initio} results: our NCSM results with the same LO, NLO and NNLOopt interactions as in Fig.~\ref{fig:1}, and the VMC calculation of Ref.~\cite{Kin20} with the NV2+3-Ia interaction. The second to last column gives the NCSM model space $N_\mathrm{max}$ while $\hbar\omega=16$~MeV for $^3$H, $^3$He and $\hbar\omega=20$~MeV for all others. The second column indicates if the transition stays within the same SU(4) irrep. 
\label{tab:2}}
\end{table*}
\endgroup

In Refs.~\cite{Pas17,Kin20} the suppression of Gamow-Teller transitions in the $A=8$ system is attributed to them connecting large and small orbital components \cite{Wir06} of nuclear states. 
In fact, since the total wave function is antisymmetric, the orbital wave function populates orbital irreps with Young tableaux that are dual to the irreps of the spin-isospin wavefunction. The spatial-structure explanation for the suppression of Gamow-Teller matrix elements adduced in Ref.~\cite{Kin20} and references therein is therefore dual to the SU(4)-content analysis carried out here (see also Ref.~\cite{Daw19}). However, analyzing spin-isospin symmetries offers some advantages compared to spatial symmetries due to the nature of the underlying groups. In a basis expansion, the spin-isospin wavefunctions populate irreps of U(4), a group of fixed dimension. In contrast, the spatial wavefunctions populate irreps of U($\Omega$), where $\Omega$ is the dimension of the corresponding single-particle basis. While this distinction does not guarantee that the spin-isospin analysis is simpler---the spin-isopsin symmetry may be more strongly broken than in the cases considered here---it generally involves fewer irreps, and those have smaller dimension. An approximate spin-isospin symmetry then constrains the allowed irreps of U($\Omega$), because of the total anti-symmetry of the wave function. This, in turn, facilitates the mapping of spatial states to relevant groups for collective dynamics, such as Elliott's SU(3) \cite{Mcc24} or the non-compact symplectic group $Sp(3,\mathbb R)$ \cite{Dyt20,Mcc20}.

%----------------------------------------
%------------CONCLUSIONS-----------------
%----------------------------------------
\textit{Conclusions}---In this letter we showed that \textit{ab initio} computed eigenstates of light nuclei have as much as 90\% of their normalization in a single SU(4) irrep. The symmetry is broken only weakly in these nuclear many-body states.
This might seem surprising in light of the large difference between the s-wave NN scattering lengths in the spin-one and spin-zero channels~\cite{Cor08}. 
However several authors have observed that, although the s-wave NN phase shifts are quite different at very low energies, they are similar for relative momenta of order 100 MeV~\cite{Tew22}. 
SU(4) invariant interactions built on this observation describe the structure of the trinucleons~\cite{Kon15,Van16} and $^4$He~\cite{Kon16} well. 
Indeed, Ref.~\cite{Kon16} argued that the relevant symmetry group for these nuclei is associated with proximity to the unitary limit~\cite{Bra06}. 
It is an important and open question whether the SU(4) symmetry emerges for that reason, because of the contracted SU(4) symmetry of hadronic interactions in the large-$N_c$ limit of QCD~\cite{Das93,Kap95,Lee21}, as a consequence of entanglement suppression~\cite{Bea18,Cav25}, or for some other reason.  

Since the Gamow-Teller operator is a generator of SU(4) it follows that Gamow-Teller transitions between nuclear states that are in different SU(4) irreps are SU(4) forbidden. However, two-body axial currents are diagonal only when restricted to two-body systems, so in such cases the effects of two-body currents is enhanced compared to the standard $\chi$EFT power counting.

In contrast, when the Gamow-Teller transition {\it is} SU(4) allowed, the result for the one-body matrix element is remarkably robust with $\chi$EFT order, changing by $< 5$\% as the order of the interaction is increased from LO to NNLO. SU(4) thus essentially guarantees that $\chi$EFT will reproduce these matrix elements to within 10\%---regardless of $\chi$EFT order---because the pure SU(4) prediction for the reduced matrix element agrees with \textit{ab initio} results at this level---a phenomenon that was already noted for tritium $\beta$-decay and $pp$ fusion in the lower-resolution pionless EFT~\cite{DeL22}. We note that the SU(4) decomposition of nuclear states is not a direct observable; a unitary transformation of the Hamiltonian could alter the weights found in this study. Such a transformation would necessarily change the balance between one- and two-body electroweak operators to ensure all nuclear observables remain constant. However, the fact that our calculated decompositions exhibit remarkable stability across four distinct $\chi$EFT interactions suggests that no simple unitary transformation exists that could simultaneously preserve the long-range component of the nuclear force and at the same time significantly modify the balance of one- to two-body operators.

These findings impact nuclear magnetic dipole (M1) transitions as well. Key parts of the M1 one-body operator can be written in terms of SU(4) generators~\cite{Nan72}. Lin et al.~\cite{Lin22} have shown that the approximate SU(4) symmetry of three-nucleon wave functions explains the otherwise surprisingly small threshold rate of $nd \rightarrow t \gamma$; Lin and Vanasse subsequently explored a dual EFT and SU(4) expansion for $\gamma t \rightarrow n d$~\cite{Lin25}. 

SU(4) symmetry is key to quantitatively explaining Gamow-Teller matrix elements, imposing patterns that sits atop any convergence in the $\chi$EFT expansion. Characterizing the SU(4) content of nuclear states is thus a prerequisite for the rigorous uncertainty quantification of $\beta$-decay amplitudes, as is needed in searches for physics beyond the Standard Model.
In future work we will use statistical tools to analyze the convergence of the $\chi$EFT series for these amplitudes~\cite{Fur15,Mel17}. We expect Wigner's SU(4) symmetry to remain relevant in heavier systems, although admixtures of different irreps might become more prominent there ~\cite{Vog93}.

%----------------------------------------
%----------ACKNOWLEDGEMENTS--------------
%----------------------------------------
\vspace{0.3cm}
\begin{acknowledgments}
We thank Garrett King for a clarifying conversation regarding the results in Table~\ref{tab:2}. This research was supported by the Swedish Research Council via a Tage Erlander Visiting Professorship, Grant No.\ 2022-00215 (D.R.P.\ and SS.L.M.), and a Research Project Grant, No.\ 2021-04507 (C.F.\ and SS.L.M.) as well as by the US Department of Energy under contract no. DE-FG02-93ER40756 (D.R.P.) and the National Science Foundation under award no. PHY-2402275. SS.L.M. acknowledges travel support from the Barbro Osher Endowement, Scholarship No.\ 90401181, the Royal Swedish Academy of Sciences, Scholarship No.\ PH2024-0090. SS.L.M. is grateful for hospitality provided by Oak Ridge National Laboratory (ORNL) and Ohio University, where part of this work was done. The computations and data handling were partly enabled by resources provided by the National Academic Infrastructure for Supercomputing in Sweden (NAISS), partially funded by the Swedish Research Council through grant agreement No.\ 2022-06725. This research used resources from the Oak Ridge Leadership Computing Facility located at ORNL, which is supported by the Office of Science of the Department of Energy under Contract No.\ DE-AC05-00OR22725. This work was supported (in part) by the U.S. Department of Energy, Office of Science, Office of Advanced Scientific Computing Research and Office of Nuclear Physics, Scientific Discovery through Advanced Computing (SciDAC) program (SciDAC-5 NUCLEI); by the U.S. Department of Energy, Office of Science.
\end{acknowledgments}

%----------------------------------------
%------------BIBLIOGRAPHY----------------
%----------------------------------------
\bibliography{./mybibfile,./betadecay}

%\begin{widetext}
\section{End Matter}
\paragraph{SU(4) structure of nuclear wavefunctions.} In the most general irreducible representations of unitary groups, state vectors can be labeled by the quantum numbers of the canonical subgroup chain $U(4) \supset U(3) \supset U(2) \supset U(1) $. Unfortunately, in this representation neither the total spin nor the isospin can be associated with quantum numbers of the chain. 
Here we follow Ref.~\cite{Hec69} and apply the subgroup chain $SU(4) \supset SU(2)_S \otimes SU(2)_T $. In this approach, we label state vectors as $\ket{[f]; S T M_S M_T; \omega \varphi}$. If the spin and isospin subgroups are contained in the $SU(4)$ irrep with multiplicity greater than one, the quantum numbers $\omega$ and $\varphi$ are required to uniquely identify spin-isospin states\footnote{This is the origin of what is sometimes referred as the \textit{state-labeling problem}.}. 
However, for most of the SU(4) irreps considered in this work\footnote{The only exception is the irrep [321], see Table \ref{tab:1} but our conclusions are unchanged even in that case.}, this multiplicity is at most one, and SU(4) states can be identified as $\ket{[f]; S T M_S M_T}$. These spin-isospin states must be coupled with orbital states $\ket{[\tilde f];\alpha L M_L}$. Here the set of space quantum numbers other than $L$ and $M_L$ has been abbreviated to $\alpha$. This orbital state has the permutation symmetry of the Young tableau $[\tilde f]$. Anti-symmetry of the total wavefunction implies that the orbital wavefunction populates the dual Young tableau $[\tilde f]$ of the spin-isospin tableau $[f]$. The states $\ket{[f]; S T M_S M_T; \alpha L M_L} \equiv \ket{[f]; S T M_S  M_T} \otimes \ket{[\tilde f]; \alpha LM_L}$ form a complete basis; Eq.~(\ref{eq: expansion}) is then the expansion of a state of good $J^2$ and $T_3$ on this basis.
The probability $P([f])$ of populating the irrep $[f]$ can be obtained as
\begin{eqnarray}
    P([f])= \sum_{\substack{T' S'M_S' \\ \alpha' L' M_L'}} |\braket{[f]; S' T' M_S' M_T' ; \alpha' L' M_L'|\Psi_{JM_T}}|^2  \ . \label{eq: probability_EM}
\end{eqnarray}
The Lanczos decomposition \cite{Joh15,Gue00} accumulates the probabilities of populating irreps with the same $\braket{C_2}$ value. Since in this work we consider only SU(4) irreps with distinct Casimir eigenvalues, see also Table \ref{tab:1}, the decomposition isolates $P([f])$. We have verified that the Lanczos algorithm reproduces Eq.~(\ref{eq: probability_EM}) by directly diagonalizing the Hamiltonian and the Casimir operator for a small NCSM model space ($N_{\mathrm{max}} =2$) in $^6$He and $^6$Li, evaluating the scalar products between the relevant eigenstates, and summing the square of the scalar products associated with identical eigenvalues of the Casimir operator. The Lanczos decomposition method and the exact diagonalization produce identical results.
\\

\paragraph{Gamow-Teller matrix elements under exact SU(4) symmetry.} Under exact SU(4) symmetry, nuclear wavefunctions have good quantum numbers $\{[f]JMLSTM_T\}$. The expansion (\ref{eq: expansion}) then simplifies to a sum over $\alpha$, $M_L$, and $M_S$:
\begin{eqnarray}
     \ket{\Psi_{\text{SU(4)}}}&=& \sum_{\substack{\alpha M_L M_S}} \braket{[f]; S T M_S M_T; \alpha LM_L|\Psi_{\text{SU(4)}}}  \nonumber \\ && \ket{[f]; S T M_S M_T; \alpha L M_L} \ .
     \label{eq: expansion_SU(4)}
\end{eqnarray}
Since the long-wavelength Gamow-Teller operator is a generator of SU(4), this decomposition allows its matrix elements to be obtained analytically in the SU(4)-symmetric limit. We write\footnote{Note that the isospin raising and lowering operators are defined as $\tau_\pm=\tau_x \pm i \tau_y$.} $GT_\pm = \sum_k \sigma_z^{(k)} \tau_\pm^{(k)}$ and obtain, by using the expansion (\ref{eq: expansion_SU(4)}) twice, 
\begin{eqnarray}
&&\braket{\Psi'_{\text{SU(4)}}|GT_\pm|\Psi_{\text{SU(4)}}}= \sum_{\substack{M_LM_S\\M_S'}} C_{SM_S,LM_L}^{JM} C_{S'M_S',LM_L}^{J'M'} \nonumber \\
&&\braket{[f];S'T'M_S'M_T'|GT_\pm|[f];STM_SM_T} \ ,
\end{eqnarray}
where the Clebsch-Gordan coefficients originate from the scalar products $\braket{[f]; S T M_S M_T; \alpha LM_L|\Psi_{\text{SU(4)}}}$ and $\braket{\Psi_{\text{SU(4)}}'|[f]; S' T' M'_S M'_T; \alpha LM_L}$. Meanwhile, the operator $GT_{\pm}$ is diagonal in $M_L$ and $\alpha$ and the sum over the quantum numbers $\alpha$ is then removed using completeness of the basis $\{\ket{\alpha L M_L}\}$ for orbital states. The last matrix element can be evaluated using group theory arguments \cite{Hec69} which yields Eq.~(\ref{eq: matrix_element_exact_su4}).
To compare our results with the recent calculation of Ref.~\cite{Kin20}, we follow their notation and define the Gamow-Teller reduced matrix element (RME) as
\begin{eqnarray}
    \text{RME}=\frac{\sqrt{2J'+1}}{2} \ \frac{\bra{\Psi'_{J'M_T'}}GT_\pm\ket{\Psi_{JM_T}}}{C_{JM,10}^{J'M}} \ , \label{eq: RME}
\end{eqnarray}
Using Eq.~(\ref{eq: RME}) and Eq.~(\ref{eq: matrix_element_exact_su4}) and the reduced Wigner coefficients from the tables in \cite{Hec69} we obtain the results shown in Table~\ref{tab:2}.
\\

\paragraph{Axial-current $\chi$EFT operator as SU(4) generator.} Up to N3LO in $\chi$EFT the two main spin-isospin structures of the two-body axial current operator can be written as linear combinations of products of two-body SU(4) generators:
\begin{eqnarray}
  (\tau^{(1)}\wedge\tau^{(2)})_a  && (\sigma^{(1)}\ \wedge \ \sigma^{(2)})_b = \nonumber \\ && 2 \ \epsilon_{a c e} \ \epsilon_{b d f} \ G^{\mathrm{2B}}_{cd} \ G^{\mathrm{2B}}_{ef} + 4 \ G^{\mathrm{2B}}_{ab} \ , \label{eq: 2b_axial_current_1} \\
   (\tau^{(1)}\wedge\tau^{(2)})_a  && \epsilon_{ijk} \ \sigma_j^{(1)} \ \sigma_l^{(2)} + (1 \rightleftharpoons 2) =  \nonumber \\ && \epsilon_{ijk} \ \epsilon_{abc} \ G^{\mathrm{2B}}_{bj} \ G^{\mathrm{2B}}_{cl} - 2 \ \delta_{il} \ G^{\mathrm{2B}}_{ka}  \ . \label{eq: 2b_axial_current_2}
\end{eqnarray}
These operators are multiplied by a function of the relative co-ordinate between the two particles, so their matrix elements between two-body SU(4)-symmetric states are straightforwardly evaluated. However, the spatial dependence of the current in the $A$-body context means the $A$-body matrix element cannot be evaluated in the same way. 
The two-body current also contains a spin-momentum structure proportional to the nucleon momentum $\mathbf{p}_i$. However, this was evaluated in Ref.~\cite{Kin20} and found to be numerically small.
\\
\paragraph{Irreducible representations of SU(4) for few-body states and tensor products}
In Table~\ref{tab:1} we identify the lowest irreps of SU(4) for $A=2$--$8$ via the total number of spin-isospin states that it contains $[\mathbf{n}]$, the value of the quadratic Casimir $\braket{C_2}$, and the Young tableau of the spin-isospin states. The dual tableau, corresponding to the symmetry of the associated spatial wavefunction is also listed.
\\
\begin{table}[h!t]
\centering
\begin{tabular}{ccccc}
\toprule
\multirow{2}{*}{Particles} & \multirow{2}{*}{[$\mathbf{n}$]} & \multirow{2}{*}{$\braket{C_2}$} & \multirow{2}{*}{Young tableau} & \multirow{2}{*}{Dual tableau} \\
& & & &\\
\hline
& & & &\\
\multirow{2}{*}{2} & $[\mathbf{6}]$ & 5 & [110] & [200]\\
& $[\mathbf{10}]$ & 9 & [200] & [110]\\
& & &\\
\multirow{3}{*}{3} & $\mathbf{[4]}$ & 3.75 & [111] & [300]\\
& $[\mathbf{20}]$ & 9.75  & [210] & [210]\\
& $[\mathbf{20}]$ & 15.75 & [300] & [111]\\
& & & &\\
\multirow{3}{*}{4} & $\mathbf{[1]}$ & 0 &[000] & [400]\\
& $\mathbf{[15]}$ & 8 & [211] & [310]\\
& $\mathbf{[20]}$ & 12 & [220] & [220]\\
& & & &\\ 
\multirow{3}{*}{6} & $[\mathbf{6}]$ & 5 & [110] & [420]\\
& $[\mathbf{10}]$& 9 &  [200] & [411]\\
& $[\mathbf{64}]$&15& [321] & [321]\\
& & &\\
\multirow{3}{*}{7} & $\mathbf{[4]}$&3.75& [111] & [430]\\
& $[\mathbf{20}]$& 9.75 &[210] & [421]\\
& $[\mathbf{36}]$& 13.75& [322] & [331]\\
& & & &\\
\multirow{3}{*}{8} & $[\mathbf{1}]$& 0 & [000] & [440]\\
& $\mathbf{[15]}$ & 8 &[211] & [431]\\
& $\mathbf{[20]}$ &12 & [220] & [422]\\
\end{tabular}
\caption{The three lowest irreps for systems made with $A=2-8$ SU(4) degrees of freedom. (Note that two-particle spin-isospin states are described by just two irreps.) The third column gives the value of the quadratic Casimir, the fourth the Young tableau of the spin-isospin wave function, and the last column the dual Young tableau.}
\label{tab:1}
\end{table}
%\\
%\paragraph{SU(4) tensor products.}
Indicating the single-nucleon quantum numbers with $\alpha\equiv\{nlm_lm_sm_t\}$, creation and annihilation operators transform under SU(4) as \cite{Hec69}
\begin{eqnarray}
    a_{\alpha_1}^\dagger &=&  [100] \ ;
    a_{\alpha_2} =  [111] \ .
\end{eqnarray}
It is possible to derive the tensor structure of general one- and two-body operators
\begin{eqnarray}
    a_{\alpha_1}^\dagger &\otimes& a_{\alpha_2} = [000] \oplus [211] \ , \\
    a_{\alpha_1}^\dagger \otimes a_{\alpha_2}^\dagger &\otimes& a_{\alpha_3} \otimes a_{\alpha_4} = 2 \times [000]\oplus 4 \times [211] \nonumber \\ 
    &\oplus& [220] \oplus [310] \oplus [422] \oplus [332] \ .
\end{eqnarray}
We will use these general expressions to derive selection rules in nuclei based on Wigner SU(4) symmetry. The trivial tensor product
\begin{eqnarray}
    [000] \otimes [f] = [f] \ ,
\end{eqnarray}
results in selection rules for nuclei with $Z=2 n$ protons and $N= 2 n$ neutrons, with $n$ any integer; under SU(4) symmetry, these systems are expected to have a dominant [000] structure and transition matrix elements are allowed if the final state has the same SU(4) structure as the transition operator. We consider next the two dominant irreps for systems with nucleon number $A=4n +2$ and maximal isospin $T=1$. From the following products
\begin{eqnarray}
     [110] \otimes [211] &=& [200] \oplus [110] \oplus [222] \oplus [321] \ , \\ {}
     [110] \otimes [220] &=& [110] \oplus [330] \oplus [321] \ , \\ {}
     [110] \otimes [310] &=& [200] \oplus [420] \oplus [411] \oplus [321] \ , \\ {} 
     [110] \otimes [422] &=& [321] \oplus [411] \oplus [433] \oplus [532] \ , \\ {}
     [110] \otimes [332] &=& [222] \oplus [321] \oplus [433] \oplus [442] \ .
\end{eqnarray}
the transition $[110] \rightarrow [200]$ is allowed only by tensors $[211]$ and $[310]$. In cases where the transition operator does not contain these structures, the transition is SU(4) forbidden. Finally, we consider systems with $A=4n\pm1$ and maximal isospin $T=3/2$. From the tensor products
\begin{eqnarray}
     [111] \otimes [211] &=& [111] \oplus [210] \oplus [322] \ , \\ {}
     [111] \otimes [220] &=& [210] \oplus [331] \ , \\ {}
     [111] \otimes [310] &=& [210] \oplus [300] \oplus [321] \ , \\ {}
     [111] \otimes [422] &=& [322] \oplus [421] \oplus [533] \ , \\ {}
     [111] \otimes [332] &=& [331] \oplus [322] \oplus [443] \ ,
\end{eqnarray}
only the tensors $[211]$, $[220]$ and $[310]$ generate transitions between the two dominant irreps~$[111]$ and $[210]$.    
%\end{widetext}

\end{document}